\documentclass[twocolumn,preprintnumbers,superscriptaddress,prl,nofootinbib,longbibliography,amsmath,amssymb]{revtex4-1}

\usepackage{graphicx}% Include figure files
\usepackage{dcolumn}% Align table columns on decimal point
\usepackage{bm}% bold math
\usepackage{color}
\usepackage{ulem}
\usepackage{gensymb}
\usepackage{braket}
\usepackage{amsmath}
\usepackage[percent]{overpic}

\begin{document}

\title{Stacking disorder and thermal transport properties of $\alpha$-RuCl$_3$}

\author{Heda Zhang}
\affiliation{Materials Science and Technology Division, Oak Ridge National Laboratory, Oak Ridge, Tennessee 37831, USA}

\author{Michael A McGuire}
\affiliation{Materials Science and Technology Division, Oak Ridge National Laboratory, Oak Ridge, Tennessee 37831, USA}

\author{Andrew F May}
\affiliation{Materials Science and Technology Division, Oak Ridge National Laboratory, Oak Ridge, Tennessee 37831, USA}

\author{Joy Chao}
\affiliation{Center for Nanophase Materials Sciences, Oak Ridge National Laboratory, Oak Ridge, Tennessee 37831, USA}

\author{Qiang Zheng}
\affiliation{Materials Science and Technology Division, Oak Ridge National Laboratory, Oak Ridge, Tennessee 37831, USA}
\affiliation{Department of Materials Science and Engineering, University of Tennessee, Knoxville, TN 37996, USA}

\author{Miaofang Chi}
\affiliation{Center for Nanophase Materials Sciences, Oak Ridge National Laboratory, Oak Ridge, Tennessee 37831, USA}

\author{Brian C Sales}
\affiliation{Materials Science and Technology Division, Oak Ridge National Laboratory, Oak Ridge, Tennessee 37831, USA}

\author{David G Mandrus}
\affiliation{Materials Science and Technology Division, Oak Ridge National Laboratory, Oak Ridge, Tennessee 37831, USA}
\affiliation{Department of Materials Science and Engineering, University of Tennessee, Knoxville, TN 37996, USA}

\author{Stephen E Nagler}
\affiliation{Neutron Scattering Division, Oak Ridge National Laboratory, Oak Ridge, Tennessee 37831, USA}

\author{Hu Miao}
\affiliation{Materials Science and Technology Division, Oak Ridge National Laboratory, Oak Ridge, Tennessee 37831, USA}

\author{Feng Ye}
\email{yef1@ornl.gov}
\affiliation{Neutron Scattering Division, Oak Ridge National Laboratory, Oak Ridge, Tennessee 37831, USA}

\author{Jiaqiang Yan}
\email{yanj@ornl.gov}
\affiliation{Materials Science and Technology Division, Oak Ridge National Laboratory, Oak Ridge, Tennessee 37831, USA}

\date{\today}

\begin{abstract}
%We report a careful investigation of the correlation between the layer stacking, structure transition, magnetic and thermal transport properties of different $\alpha$-RuCl$_3$ crystals with T$_N$ varying from 6\,K to 7.6\,K by measuring magnetic properties, specific heat, neutron single crystal diffraction, and thermal transport. The stacking disorder induces the variation in T$_N$. $\alpha$-RuCl$_3$ crystals with a single T$_N$ in the range of 7.4\,K-7.6 K have minimal amount of stacking disorder and show a well defined structure transition around 140\,K upon cooling.  For those crystals with more stacking faults and a T$_N$ below 7\,K, the structure transition occurs well below 140\,K upon cooling and is incomplete, manifested by the coexistence of both high temperature and low temperature phases down to the lowest measurement temperature. Diffuse streaks were observed for these crystals but not for those with T$_N$ above 7.4\,K. For both types of crystals, oscillatory field dependent thermal conductivity and plateau-like feature in thermal Hall resistivity were observed in the field-induced disordered state that may be a quantum spin liquid. However, $\alpha$-RuCl$_3$ crystals with minimal amount of stacking disorder have a higher thermal conductivity that pushes the thermal Hall conductivity to be close to the half-integer quantized value. The strong correlation between the layer stacking, structure transition, magnetic and thermal transport properties highlights the importance of interlayer coupling in $\alpha$-RuCl$_3$ despite the weak van der Waals bonding.

$\alpha$-RuCl$_3$, a well-known candidate material for Kitaev quantum spin liquid, is prone to stacking disorder due to the weak van der Waals bonding between the honeycomb layers. After a decade of intensive experimental and theoretical studies, the detailed correlation between stacking degree of freedom, structure transition, magnetic and thermal transport properties remains unresolved. In this work, we reveal the effects of a small amount of stacking disorder inherent even in high quality $\alpha$-RuCl$_3$ crystals. This small amount of stacking disorder results in the variation of the magnetic ordering temperature, suppresses the structure transition and thermal conductivity. Crystals with minimal amount of stacking disorder have a T$_N>$7.4\,K and exhibit a well-defined structure transition around 140\,K upon cooling. For those with more stacking faults and a T$_N$ below 7\,K, the structure transition occurs well below 140\,K upon cooling and is incomplete, manifested by the diffuse streaks and the coexistence of both high temperature and low temperature phases down to the lowest measurement temperature. Both types of crystals exhibit oscillatory field dependent thermal conductivity and a plateau-like feature in thermal Hall resistivity in the field-induced quantum spin liquid state. However, $\alpha$-RuCl$_3$ crystals with minimal amount of stacking disorder have a higher thermal conductivity that pushes the thermal Hall conductivity to be closer to the half-integer quantized value. These findings demonstrate a strong correlation between layer stacking, structure transition, magnetic and thermal transport properties, underscoring the importance of interlayer coupling in $\alpha$-RuCl$_3$ despite the weak van der Waals bonding.

 \end{abstract}

\maketitle

Shortly after the first experimental investigation of $\alpha$-RuCl$_3$ as the candidate material for Kitaev quantum spin liquid\cite{plumb2014alpha}, it was realized that the stacking sequence of the RuCl$_3$ honeycomb layers can affect the magnetic properties \cite{banerjee2016proximate}. This is typical for van der Waals bonded cleavable magnets where nowadays the stacking degree of freedom of the two dimensional structure units has been employed to engineer magnetic ground states~\cite{huang2017layer,chen2019direct,yan2022perspective}. The stacking of RuCl$_3$ honeycomb layers relative to each other leads to different proposed crystal structures including the monoclinic \textit{C}2/\textit{m}, the rhombohedral $R\bar{3}$, and the trigonal $P3_112$. At room temperature, most studies reported a $C2/m$ structure for $\alpha$-RuCl$_3$. Upon cooling, this monoclinic structure becomes unstable and around 150\,K $\alpha$-RuCl$_3$ goes through a first order structure transition most likely to the rhombohedral $R\bar{3}$ structure~\cite{mu2022role}. In early studies of $\alpha$-RuCl$_3$, more than one magnetic anomaly is typically observed in the temperature range 7-14\,K~\cite{kubota2015successive,johnson2015monoclinic,sears2015magnetic,majumder2015anisotropic,banerjee2016proximate}. The 14\,K magnetic order seems to result from the $C2/m$ type stacking~\cite{cao2016low,johnson2015monoclinic}. And it is believed that the mixing of different types of stacking gives rise to the multiple magnetic anomalies in the temperature range 7\,K-14\,K.

With the intense materials synthesis effort, most $\alpha$-RuCl$_3$ crystals being studied nowadays have only one single magnetic transition around T$_N$=7\,K. This has been employed as a simple and convenient criteria for a quick check of crystal quality. However, T$_N$ has been found to vary from 6\,K to 8\,K~\cite{kim2022alpha,sears2017phase,do2017majorana}. The origin of this discrepancy is still unknown. Also unknown is whether and how this magnetic order is correlated with the first order structure transition at high temperatures.

Recently, the intensively debated thermal transport properties of $\alpha$-RuCl$_3$ calls for a thorough revisit to the materials issues of this fascinating compound~\cite{lee2021}. The half-integer quantized thermal Hall conductance was believed to be one of the fingerprints for Majorana fermions of the fractionalized spin excitations in $\alpha$-RuCl$_3$. However, the thermal Hall conductance in the field-induced disordered or quantum spin liquid state was found to be sample dependent and the experimental observation of half-integer quantized value requires using $\alpha$-RuCl$_3$ crystals with high longitudinal thermal conductivity~\cite{yokoi2021half,kasahara2018majorana,yamashita2020sample,lefranccois2022evidence,czajka2022planar,bruin2022robustness,kasahara2022quantized}. The other intriguing experimental observation is the oscillatory features of thermal conductivity as a function of in-plane magnetic field. These features were reproduced by different groups but the origin is under hot debate. While some believed this is an intrinsic character of the magnetic phase of T$_N\approx$7\,K~\cite{Zhang14K,czajka2021oscillations} and attributed the observed oscillations to quantum oscillations of putative charge-neutral fermions~\cite{czajka2021oscillations}, others believed that the oscillatory features have an extrinsic origin and result from a sequence of field-induced magnetic phase transitions in crystals with stacking disorder~\cite{bruin2022origin,lefranccois2023oscillations}.

All these interesting but debated results and interpretations highlight the importance of understanding the materials issues in $\alpha$-RuCl$_3$. In this work, we performed a careful investigation of the correlation between the layer stacking, structure transition, magnetic and thermal transport properties of different $\alpha$-RuCl$_3$ crystals with T$_N$ varying from 6\,K to 7.6\,K. The amount of stacking disorder discussed in this work is far less than that in previous studies~\cite{kubota2015successive,johnson2015monoclinic,sears2015magnetic,majumder2015anisotropic,banerjee2016proximate} and may not induce significant magnetic anomalies at 10-14\,K in magnetic and specific heat measurements. This small amount of stacking disorder causes the variations in T$_N$ in different $\alpha$-RuCl$_3$ crystals. Based on the characterizations, we categorize our $\alpha$-RuCl$_3$ crystals into two types. Type-I $\alpha$-RuCl$_3$ single crystals with T$_N$ above 7.2\,K have minimal amount of stacking disorder and show a well defined structure transition around 140\,K upon cooling. For type-II crystals with more stacking disorder and a lower T$_N$, the structure transition occurs below 140\,K upon cooling and is incomplete, manifested by the coexistence of both high temperature and low temperature phases and the observation of diffuse streaks by neutron scattering below the structure transition. For both types of crystals, oscillatory field dependent thermal conductivity and plateau like feature in thermal hall resistivity were observed in the field-induced quantum spin liquid state. However, $\alpha$-RuCl$_3$ single crystals with T$_N$ above 7.4\,K have a higher thermal conductivity, which pushes the thermal Hall conductivity to be closer to the half integer quantized value. Our results demonstrate a strong correlation between the layer stacking, structure transition, magnetic and thermal transport properties and highlight the importance of interlayer coupling in $\alpha$-RuCl$_3$.

\section{Experimental details}
$\alpha$-RuCl$_3$ used in this study were grown using two vapor transport techniques:  self-selecting vapor growth with a small temperature gradient near the powder, and the conventional vapor transport technique with a large temperature gradient. The former is employed to grow thicker crystals for neutron single crystal diffraction measurements and the growth details can be found elsewhere~\cite{yan2022self}. The latter was performed in a two-zone tube furnace with the hot end kept at 1000$\degree$C and the cold end at 750$\degree$C. This yields millimeter sized single crystals with minimal amount of stacking disorder that are ideal for measurements of magnetic, thermodynamic, and thermal transport properties. The growth details are reported previously\cite{Zhang14K}. Varying the growth temperature and temperature gradient along the growth ampoule impacts the layer stacking and thus physical properties. For example, $\alpha$-RuCl$_3$ crystals with T$_N$ around 6\,K were obtained by keeping the hot end of the growth ampoule at 1075$\degree$C while the cold end at 1025$\degree$C. We noticed that crystal quality and properties are sensitive to the total vapor pressure inside of the growth ampoule. Our growths suggest a  higher vapor pressure facilitates the formation of stacking disorder, although a detailed relation is not developed yet. Therefore, in order to obtain $\alpha$-RuCl$_3$ crystals with controlled degree of stacking disorder, all factors affecting the total vapor pressure inside of the growth ampoule should be considered, such as, the purity and amount of starting powder, the volume of growth ampoule, growth temperature and temperature gradient along the growth ampoule. In our growths, high pure RuCl$_3$ powder synthesized by reacting RuO$_2$ powder with Al$_2$O$_3$-KCl salt\cite{yan2017flux} or purchased from Furuya Metals (Japan) was used and they are found to be comparable with respect to disorder control.

Elemental analysis on cleaved surfaces was performed using the wavelength dispersive (WDS) spectroscopy techniques. WDS measurement was performed using a JEOL JXA-8200X electron microprobe analyzer instrument equipped with five crystal-focusing spectrometers for wavelength dispersive x-ray spectroscopy. Magnetic properties were measured with a Quantum Design (QD) Magnetic Property Measurement System in the temperature range 2.0\,K$\leq$T$\leq$\,300\,K. Specific heat data below 30\,K were collected using a QD Physical Property Measurement System (PPMS).  Thermal transport measurements were performed on a custom-built PPMS puck as described before\cite{Zhang14K}. The temperature dependence of 0 0 \textit{l} reflections was monitored on a flat surface by a PANalytical X’Pert Pro MPD powder x-ray diffractometer using Cu K$_{\alpha1}$ radiation. An Oxford PheniX closed cycle cryostat was used to measure from 300\,K to 20\,K.

Neutron diffraction experiments were carried out using the single crystal diffractometer CORELLI~\cite{ye18} at SNS to monitor the structural phase  transition and to search for potential diffuse scattering arising from possible stacking disorder in crystals grown under different conditions. Single crystals are mounted inside
a closed-cycle refrigerator with a base temperature of 6\,K. The Mantidworkbench package was used for data reduction and analysis. We define the momentum transfer Q in $3D$ reciprocal space in $\AA^{-1}$ as $Q= Ha^* + Kb^* + Lc^*$, in which $H$, $K$ and $L$ are Miller indices, and $a^*$, $b^*$, $c^*$ are the lattice vectors in reciprocal space. For each type of crystals mentioned below, we checked one small piece around 6\,mg and one large piece around 200\,mg. No difference was observed between the small and large pieces.

\section{Results}

\begin{figure*} \centering \includegraphics [width = 0.95\textwidth] {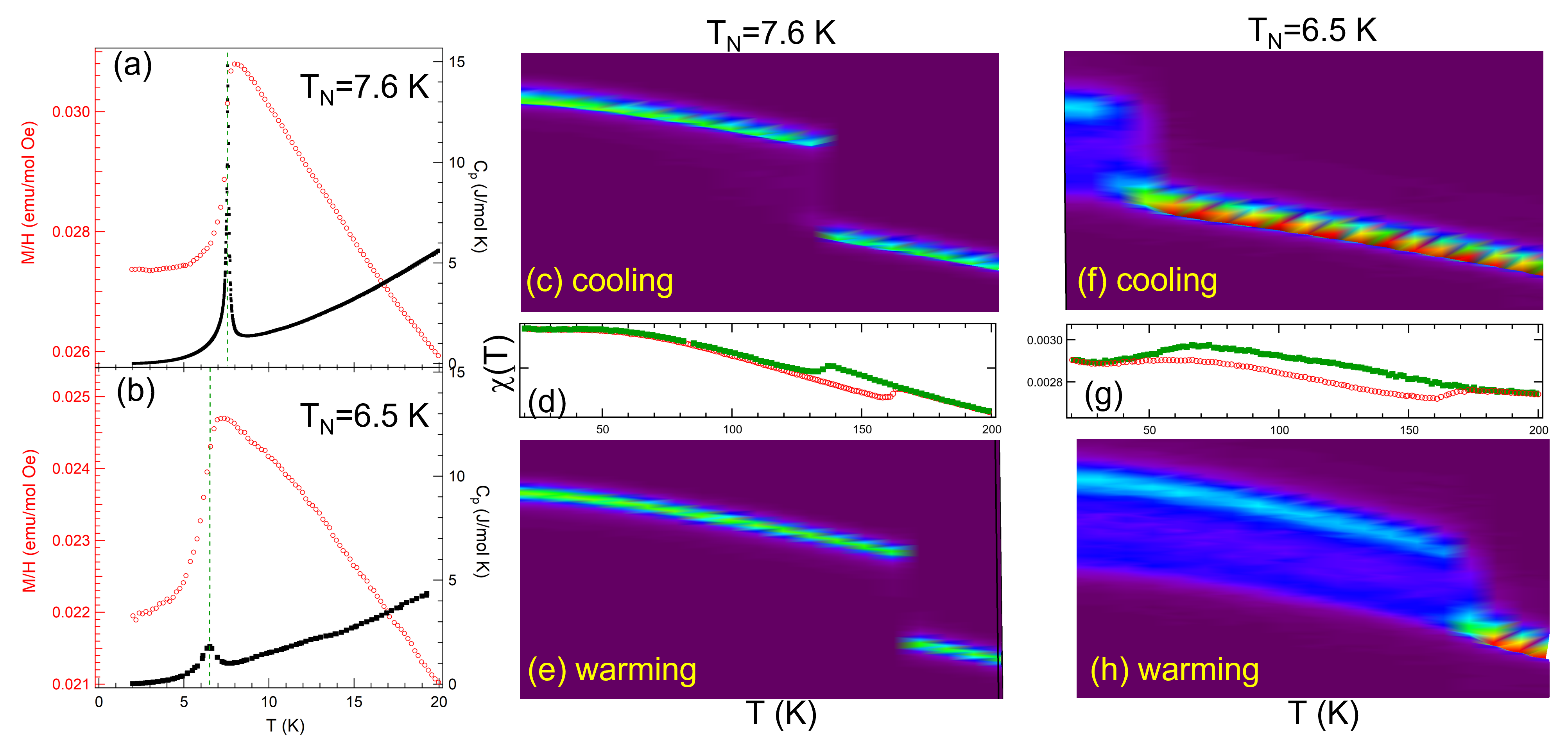}
\caption{(color online) Magnetic order and structure transition in two typical types of $\alpha$-RuCl$_3$ single crystals. As described in the text, type-I crystals have magnetic ordering temperatures above 7.2\,K, while type-II crystals normally order magnetically below 7\,K. (a,b) Temperature dependence of magnetization and specific heat (Cp) below 20\,K. The magnetization was measured in a magnetic field of 1\,kOe applied along the zig-zag direction (perpendicular to the Ru-Ru bond). The vertical dashed lines highlight the Neel temperature, T$_N$, defined as the temperature where Cp peaks. T$_N$=7.6\,K in (a) is 1.1\,K higher than that in (b).  (c-e) Temperature dependence of magnetization and 005 reflection in the temperature range 20\,K-200\,K for the crystal with T$_N$=7.6\,K, a representative for type-I crystals. (f-h) Temperature dependence of magnetization and 005 reflection in the temperature range 20\,K-200\,K for the crystal with T$_N$=6.5\,K, an example for type-II crystals.  The temperature dependence of magnetization in (d,g) was measured in a magnetic field of 10\,kOe applied perpendicular to the honeycomb plane.}
\label{MagCpXRD-1}
\end{figure*}

Figure\,\ref{MagCpXRD-1} shows the temperature dependence of magnetization, specific heat, and intensity of the 005 structural Bragg reflection (in \textit{C}2/\textit{m} notation) illustrating the magnetic ordering and structure transition temperatures for two typical types of single crystals. We define the magnetic ordering temperature as the temperature where Cp(T) peaks. Type-I crystals have T$_N>$7.2\,K and show sharp anomalies at T$_N$ in both temperature dependence of magnetization and specific heat. Results of one crystal with T$_N$=7.6\,K (see Fig.\,\ref{MagCpXRD-1}a) are presented in this work as a representative of type-I crystals. For this crystal, abrupt changes in the temperature dependence of magnetization (Fig.\,\ref{MagCpXRD-1}d) and 005 reflection suggest the structure transition occurs at about 140\,K upon cooling (Fig.\,\ref{MagCpXRD-1}c) and about 170\,K upon warming(Fig.\,\ref{MagCpXRD-1}e). Upon cooling below 140\,K, the intensity of the peak sitting at 2$\theta$\,=\,85.15\,degree decreases quickly upon further cooling and disappears around 130\,K. Meanwhile, a peak sitting at 2$\theta$\,=\,85.62\,degree appears and its intensity increases quickly upon cooling and saturates below about 120\,K. Step-like features associated with the structure transition are observed in the high temperature magnetization at 140\,K and 170\,K upon cooling and warming, respectively. Similar features are observed for all $\alpha$-RuCl$_3$ crystals with T$_N$ above 7.2\,K.

Type-II crystals (Fig.\,\ref{MagCpXRD-1}b) normally have a T$_N$ below 7\,K and show broader transitions and weaker anomalies near T$_N$ in magnetization and specific heat. Physical properties of one crystal with T$_N$=6.5\,K were presented in this work as an example of type-II crystals. Above T$_N$, some weak features were observed in both magnetization and specific heat in (b), in sharp contrast to (a). The structure transition also shows anomalous features distinct from those for type-I crystals. Upon cooling, the structure transition of the crystal with T$_N$=6.5\,K (see Fig.\,\ref{MagCpXRD-1}f) occurs around 60\,K and the transition is not complete which is well illustrated by the coexistence of reflections from both high temperature and low temperature phases to the lowest temperature 20\,K of our x-ray diffraction measurement. Upon warming from 20\,K, the phase coexistence persists until about 170\,K above which only the high temperature monoclinic phase remains(Fig.\,\ref{MagCpXRD-1}h). Accordingly, a wide loop was observed in the temperature dependence of magnetization (Fig.\,\ref{MagCpXRD-1}g). It is interesting to note that for both $\alpha$-RuCl$_3$ crystals the structure transition occurs around 170\,K upon warming. When screening crystals, we noticed that the structure transition can occur in a wide temperature range 50\,K-140\,K when measuring upon cooling. For type-II crystals, the phase coexistence is always observed for crystals with T$_N$ below 7\,K.  Fig.S1 of Supporting Materials shows the x-ray diffraction results of two different crystals. Both crystals have T$_N$=7.1\,K (from specific heat) and the structure transition at 120\,K upon cooling. However, one crystal shows a complete structure transition while the other one shows phase coexistence below 120\,K. 120\,K seems to be the lowest structure transition temperature that one $\alpha$-RuCl$_3$ crystal can have before showing a sluggish first order structure transition and phase coexistence at low temperatures.

\begin{figure*} \centering \includegraphics [width = 0.95\textwidth] {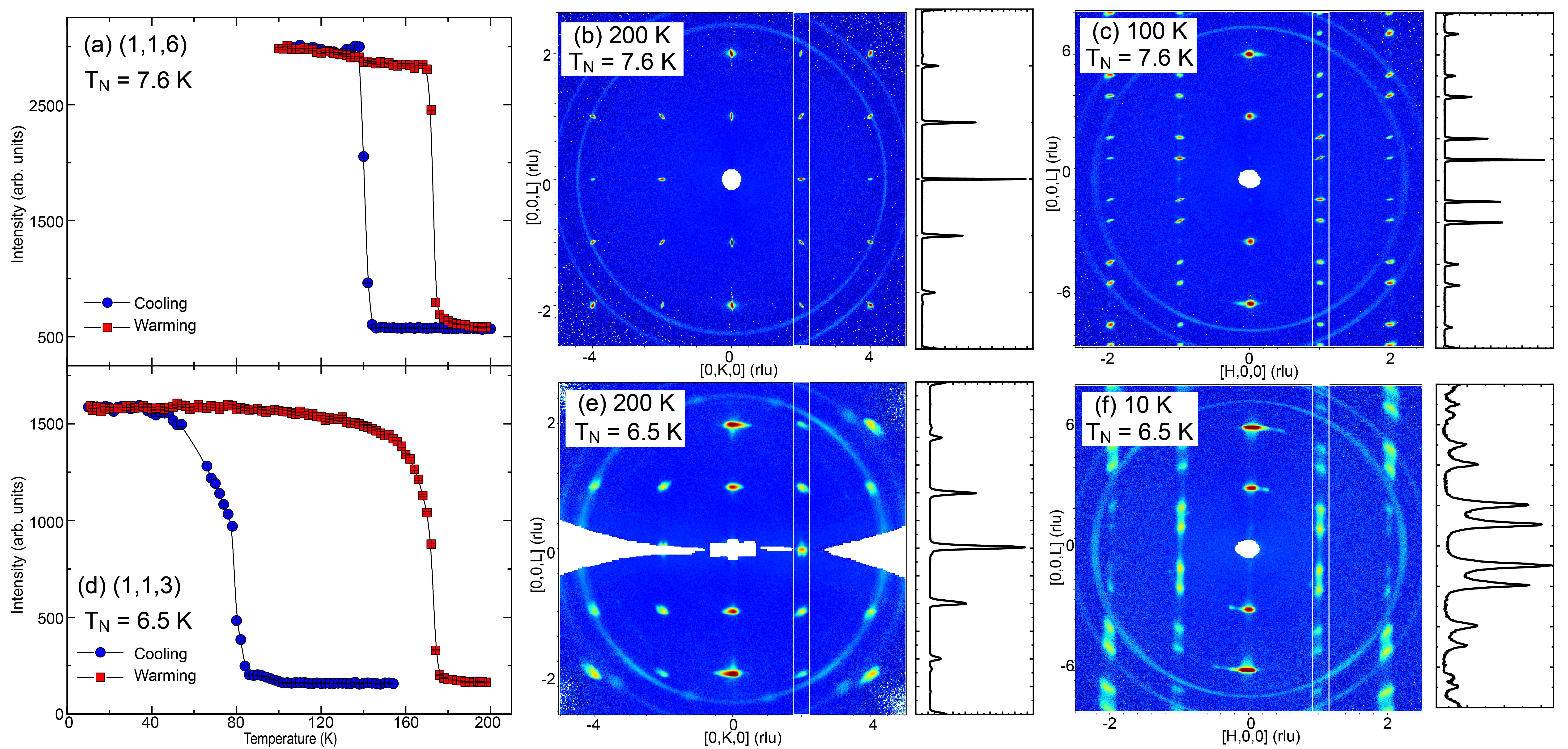}
\caption{(color online) Neutron single crystal diffraction found a sharp structure transition for the crystal with T$_N$=7.6\,K but a sluggish structure transition and diffuse streaks for the crystal with T$_N$=6.5\,K. (a) The thermal evolution of the $(1,1,6)$ reflection measured in cooling and warming confirms the
sharp structural transition in the crystal with T$_N$=7.6\,K. (b, c) The neutron diffraction patterns at 200~K
(in the high temperature $C/2m$ phase) and 100~K (in the low temperature $R\bar{3}$ phase) for the crystal with T$_N$=7.6\,K. Extra reflections in
(c) confirm the symmetry change at low temperatures. (d) The temperature evolution of $(1,1,3)$ reflection
measured in cooling and warming of the crystal with T$_N$=6.5\,K.
(e, f) The neutron diffraction patterns at 200~K (in the high temperature $C/2m$ phase) and 10~K (in the low temperature $R\bar{3}$ phase) for the crystal with T$_N$=6.5\,K. The mosaic in (e) and the diffuse streaks in (f) are highlighted by the line cut shown in the right half of each panel (b,c,e,f). The transformation matrix between the
high-$T$ $C2/m$ and low-$T$ $R{\bar3}$ unit cell are: ${\bf a}_r = {\bf a}_m$,
${\bf b}_r = 1/2(-{\bf a}_m+{\bf b}_m)$, ${\bf c}_r = {\bf a}_m + 3 {\bf c}_m$,
where the subscript ``r'' and ``m'' denote the rhombhedral and monoclinic structures.
}
\label{Neutron-1}
\end{figure*}

Considering the limited penetration depth of x-ray diffraction, we studied the structure of different pieces of pristine $\alpha$-RuCl$_3$ crystals using single crystal neutron diffraction. All crystals studied with neutron diffraction were well characterized by measuring magnetic properties and/or specific heat before exposing to neutron beam. Multiple pieces with mass ranging from 6\,mg to 200\,mg and T$_N$ ranging from 6\,K to 7.6\,K were checked. All crystals crystallize in \textit{C}2/\textit{m} at room temperature but have different degree of mosaic. We are aware of previous reports of successful growth of $\alpha$-RuCl$_3$ crystals with the $P3_112$ structure~\cite{bruin2022origin}. Our neutron results do not rule out the possibility that some small crystals suitable for x-ray single crystal diffraction can have $P3_112$ type structure. Figures\,\ref{Neutron-1}(a,b) show the temperature dependence of (116) or (113) reflection to determine the structure transition in both crystals. For the crystal with T$_N$=7.6\,K, the structure transition occurs around 140\,K upon cooling and 170\,K upon warming, consistent with those determined from magnetic and x-ray diffraction measurements shown in Fig.\,\ref{MagCpXRD-1}.  For the crystal with T$_N$=6.5\,K, the structure transition occurs below 80\,K upon cooling and 170\,K upon warming. The latter temperature agrees well with those determined from magnetic and X-ray diffraction measurements shown in Fig.\,\ref{MagCpXRD-1}.  The slightly different transition temperature when measuring cooling is in line with the fact that this transition temperature can vary in a wide temperature range 50\,K-140\,K depending on the concentration and detailed distribution of stacking imperfection.

\begin{figure*} \centering \includegraphics [width = 0.9\textwidth] {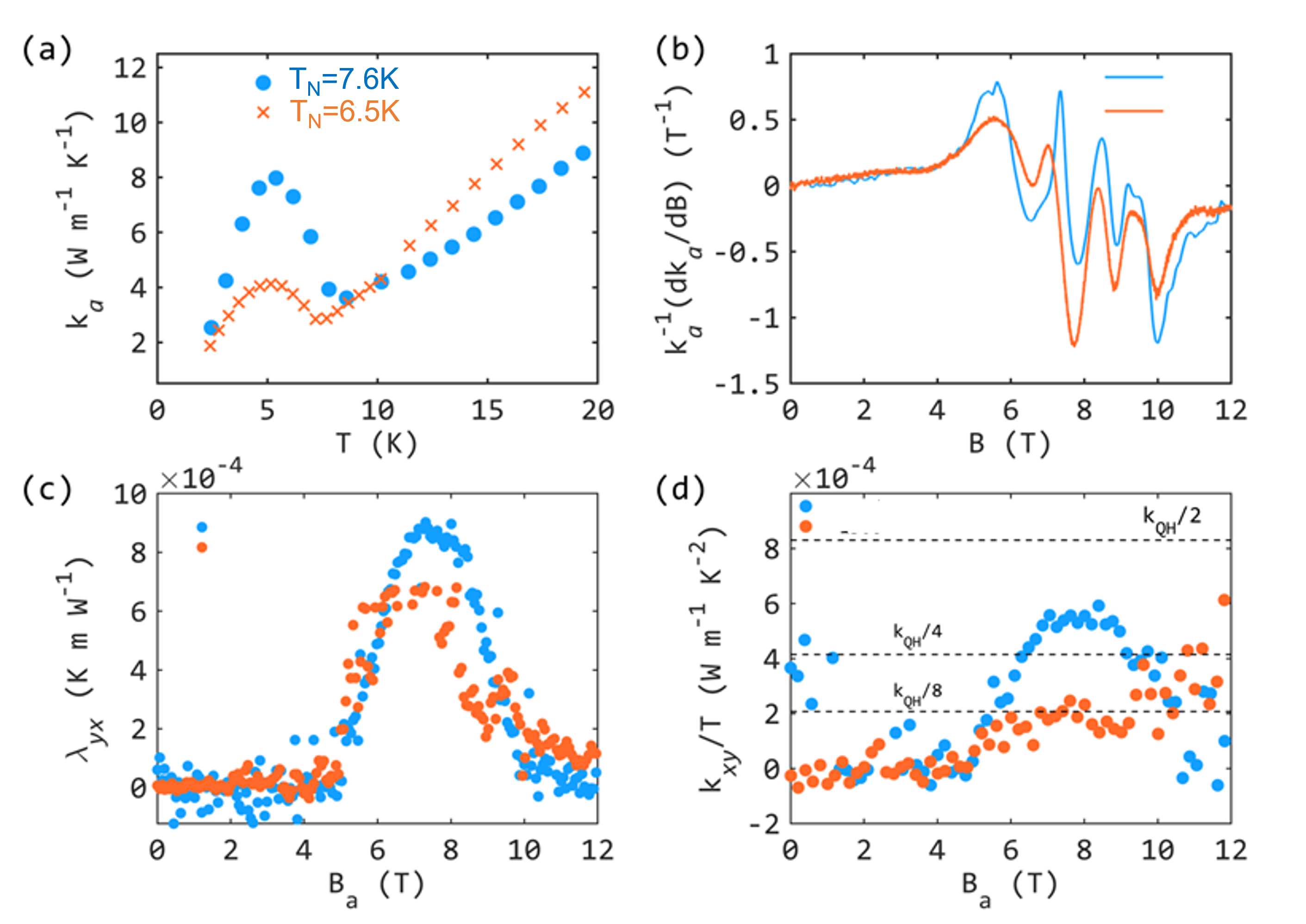}
\caption{(color online) Thermal transport measurement results of $\alpha$-RuCl$_3$. (a) Thermal conductivity of $\alpha$-RuCl$_3$ crystal. Heat current is applied along the zig-zag direction (perpendicular to the Ru-Ru bond). (b) Oscillatory features of thermal conductivity. Data collected with both heat current and magnetic field along the zig-zag direction. (c, d) Thermal Hall resistivity (c) and conductivity (d) at 5\,K. Both heat current and  magnetic field are applied along the zig-zag direction. The horizontal dashed lines in (d) highlight the thermal Hall conductivity relative to the fraction of quantized thermal Hall $\kappa_{QH}$/n, where n\,=\,2,4,and 8 are shown. }
\label{Kappa-1}
\end{figure*}

Detailed reciprocal space maps provide better understanding of the nature of average and local structure for both $\alpha$-RuCl$_3$ single crystals.
Figures~\ref{Neutron-1}b and c show the neutron diffraction patterns collected at 200~K (above the structure transition) and 100~K (below the structure transition) for the type-I crystal with T$_N$=7.6\,K. As reported previously, the high temperature monoclinic $C2/m$ structure becomes unstable upon cooling and transforms into a rhombohedral $R\bar{3}$ structure~\cite{mu2022role}. This can be appreciated by the appearance of  additional Bragg peaks along the $L$ direction (see Fig.\,\ref{Neutron-1}c). The right half of each panel shows the line cut along $[0,2,L]$ highlighting that  (1) this crystal is of high quality with negligible mosaic and (2) the absence of any diffuse feature in the temperature range investigated. The correlation length $\xi$ derived from the full width half maximum (FWHM) of the $(0,2,0)$ reflection at 200~K using Lorentzian line-shape is $\rm \sim 450\pm10~\AA$, which is at the resolution limit of the neutron measurement. The peak broadens slightly at 100~K and suggests a reduction of the correlation length to about $\rm 350\pm10~\AA$. Figures~\ref{Neutron-1}(e)-\ref{Neutron-1}(f) show the neutron diffraction patterns at 200~K (above the structure transition) and 10~K (below the structure transition) for the type-II crystal with T$_N$=6.5\,K. The crystal clearly has a much larger mosaic and more interestingly, diffuse streaks are evident below the structural transition [Fig.~\ref{Neutron-1}(f)].  The correlation length is about $\rm 78\pm5~\AA$ at 200~K and reduces to $\rm 32\pm 3~\AA$ at 10~K. Both values are much smaller compared to those for the crystal with T$_N$=7.6\,K. Despite the apparent difference in the correlation lengths, the local environment surrounding the Ru ions for both single crystals are very similar in the high temperature $C2/m$ phase. As  can be determined from the structure information listed in Table I in Supporting Materials, those two crystals have the same lattice parameters, Cl-Ru-Cl bond angles, Ru-Cl bond lengths at 200\,K. Given the surprisingly similar crystal structure and lack of any unidentified phase, the observed diffuse streaks in the crystal with T$_N$=6.5\,K are attributed to the stacking faults that are prevalent in van der Waals bonded layered materials. As discussed later, the stacking disorder can act as the pinning center preventing a uniform structure phase transition. Thus the structure transition occurs at a lower temperature. The inhomogeneous distribution of stacking disorder might be responsible for the sluggish transition occurring in a wider temperature range.

After warming above the structure transition, both crystals seem to get back to the original pristine state. In Supporting Materials, we show the neutron diffraction patterns of both crystals after warming up to 200\,K from below the structure transition. Interestingly, the diffraction pattern and the correlation length are comparable before and after the thermal cycling. We also monitored possible effects of thermal cycling on the structure transition and magnetic order by measuring the temperature dependence of magnetization and specific heat. As shown in the Supporting Materials, the thermal cycling does not seem to affect the magnetic order after 10 thermal cycles.

Figure\,\ref{Kappa-1}(a) shows the typical temperature dependence of thermal conductivity ($\kappa_a$, thermal current along \textit{a}-axis) of both single crystals below 20\,K. The small difference in T$_N$ can be well resolved. Both samples show a recovery of heat transport upon cooling through T$_N$ and a peak around 5\,K in the temperature dependence of thermal conductivity. This is consistent with previous reports\cite{hentrich2018unusual, hirobe2017magnetic,leahy2017anomalous,kasahara2022quantized,lefranccois2022evidence}. The crystal with T$_N$=7.6\,K has a higher thermal conductivity below T$_N$. This magnitude of thermal conductivity is higher than the threshold value for the observation of half-integer quantized thermal Hall effect~\cite{kasahara2022quantized}.

Figure\,\ref{Kappa-1}(b) shows the oscillatory features in the field dependence of longitudinal thermal conductivity ($\kappa_a$) at 2\,K, shown as the scaled derivative ($\kappa^{-1}$d$\kappa$/dB T$^{-1}$) of thermal conductivity with respect to applied magnetic field. This scaling procedure brings the oscillatory features of various samples with different thermal conductivities to the same scale and emphasizes the critical fields instead of absolute oscillatory amplitudes. We observed that the oscillatory features, especially those after the system has entered the field-induced quantum spin liquid phase (~7.5 T), are very similar for both samples. This indicates that the oscillatory features in the field induced quantum spin liquid region are dominated by in-plane physics.

We also studied the thermal Hall effect of both crystals. Figures.\,\ref{Kappa-1}c and d show a plateau region in both thermal resistivity and (scaled) conductivity measured at 5\,K. The plateau spans the field region in which the field-induced disordered or quantum spin liquid phase dominates. The profile of the thermal Hall signal is better shown by resistivity. The scaled thermal Hall conductivity is highly sample dependent. We noticed in our sample screening that the plateau value could vary by a factor of 5 for different samples with different thermal conductivities. The crystal with T$_N$=7.6\,K has less stacking disorder and hence a higher thermal conductivity. This leads to a near half-integer quantized $\kappa_{xy}$/T. In contrast, $\kappa_{xy}$/T for the crystal with T$_N$=6.5\,K is further away from the half-integer quantized value as illustrated by the data shown in Fig.\,\ref{Kappa-1}(d). A more detailed study of the anisotropic thermal transport properties and the origin of the oscillatory features will be reported separately.

\section{Discussions}

The detailed physical properties of two $\alpha$-RuCl$_3$ crystals were reported in this work as the representatives of two types of $\alpha$-RuCl$_3$ crystals. Typically, type-I crystals have a T$_N$ above 7.2\,K, whereas type-II crystals have a T$_N$ below 7\,K.  Specific heat results shown in Fig.\,\ref{MagCpXRD-1} and Fig. S5 in Supporting Materials demonstrate that T$_N$ of our $\alpha$-RuCl$_3$ crystals can range from 6\,K to 7.6\,K. Our WDS measurements confirmed that all crystals have the same atomic ratio and the \textit{nonstoichiometry or chemical composition should not be responsible for the variation in T$_N$ among different $\alpha$-RuCl$_3$ crystals}. Therefore, the differences in properties between type-I and type-II crystals may shed some light on the origin of T$_N$ variation in different $\alpha$-RuCl$_3$ crystals.

A careful comparison between type-I and -II crystals indicates that \textit{stacking disorder causes a sluggish structure transition and leads to phase coexistence at low temperatures}. Type-II crystals distinguish themselves from type-I crystals by showing a sluggish structure transition most likely originating from the stacking disorder. The larger mosaic (see Fig.\,\ref{Neutron-1}e) suggests the presence of stacking disorder in type-II crystals. From magnetic and x-ray and neutron diffraction measurements of different crystals, the structure transition when being measured upon cooling occurs at a lower temperature with increasing mosaic, and two-phase coexistence is always observed when the structure transition occurs below 120\,K. The phase coexistence was observed at 20\,K, the lowest temperature of our X-ray diffraction measurements. It will not be a surprise if the phase coexistence persists at even lower temperatures, for example, below T$_N$. Our x-ray and neutron diffraction studies provide solid evidence for the phase coexistence but cannot tell the detailed concentration and distribution of those two phases or the stacking disorder. However, one can expect the stacking disorder acts as a pinning center for the sliding of RuCl$_3$ honeycomb layers during the structure phase transition, leading to more complex stacking disorder as the temperature decreases. This expectation is consist with the observation of diffuse streaks for type-II crystals shown in Fig.\,\ref{Neutron-1}(f). Such diffuse streaks were observed in previous studies and attributed to stacking disorder between RuCl$_3$ honeycomb layers~\cite{kim2022alpha,johnson2015monoclinic}.
The concentration and detailed distribution of stacking disorder in the starting crystals affect the mosaic, structure transition temperature, the concentration and distribution of both high temperature and low temperature phases below the structure transition. Figure S1 presents the x-ray diffraction results for crystals with T$_N$ near 7.1\,K, which highlights the effect of the concentration and detailed distribution of stacking disorder.

The neutron single crystal diffraction confirms that type-I crystals are of high quality as evidenced by a large correlation length beyond the resolution limit of our neutron measurement. However, this doesn't mean that these crystals are free of stacking disorder. Previously, a scanning transmission electron microscope (STEM) study observed $P3_112$ type stacking in $\alpha$-RuCl$_3$ flakes from one crystal with $C2/m$ structure determined by neutron single crystal diffraction~\cite{ziatdinov2016atomic}. This result was reproduced in our recent STEM studies. STEM probes a small region of a thin flake, while neutron diffraction studies the whole crystal in neutron beam averaging over many similar and different regions. The discrepancy between STEM and neutron diffraction indicates the presence of stacking disorder in type-I crystals and the disorder can even exist in local regions in the form of other types of layer stacking with comparable energy\cite{cao2016low}. Nevertheless, the long coherence length implies the concentration of stacking disorder in high quality type-I crystals is low.

\textit{The stacking disorder can be responsible for the slightly suppressed T$_N$}. The phase coexistence or stacking disorder may affect both the interlayer and intralayer magnetic interactions. The interlayer magnetic interactions are disrupted by inhomogeneous layer spacing and exchange paths, while theoretical studies have shown that the intralayer exchange interactions are influenced by the stacking structure~\cite{kim2016crystal}. Neutron single crystal diffraction at 200\,K revealed the same local structure in both types of $\alpha$-RuCl$_3$ crystals. However, whether the local structure similarity persists at low temperatures is unknown. Despite tremendous efforts, the low temperature structure is not yet determined due to the appearance of diffuse scattering below the structure transition\cite{cao2016low,johnson2015monoclinic}. Solving the low temperature structure is beyond the scope of current effort. However, we would point out that this is essential to resolve possible local distortion of RuCl$_6$ octahedra that determines the relative magnitude of those terms in magnetic Hamiltonian and also critical for a quantitative understanding of the suppression of T$_N$. In addition, the phase coexistence scenario should be considered when solving the low temperature structure, depending on the quality of crystals employed.

The above discussion of how stacking disorder suppresses T$_N$ does not contradict the previous proposal that mechanical deformation can induce magnetic anomalies in the temperature range 10-14\,K\cite{cao2016low,Zhang14K}. In previous studies, mechanical deformation was employed to create a large concentration of stacking disorder, which resulted in crystals with one single magnetic order at 14\,K. However, the concentration of stacking disorder in both types of crystals studied in this work is much lower and this is especially true to type-I crystals. On the other hand, the suppression of T$_N$ from 7\,K and the appearance of weak magnetic anomalies in the temperature range 10-14\,K seem to be strongly correlated and may have the same origin. The Supporting Materials present specific heat data for more samples, with separate panels highlighting the features below and above 8\,K. For those crystals with suppressed T$_N$ (for example lower than 7\,K) or with more than one lambda type anomaly, some weak anomalies can be observed above 8\,K. All the observations suggest that a small amount of stacking disorder suppresses the long range magnetic order with T$_N$ around 7\,K, and increasing the population of stacking faults leads to magnetic phases with T$_N$ above 10\,K.

\textit{The stacking disorder is also responsible for the sample dependent thermal transport properties}. Upon cooling through T$_N$, type-I crystals show a stronger recovery of thermal conductivity (see Fig.\,\ref{Kappa-1}a). The high thermal conductivity and T$_N$ of type-I crystals indicate the importance of interlayer coupling and interactions. As reported previously~\cite{kasahara2022quantized,yamashita2020sample}, a large thermal conductivity is necessary for the observation of half-integer quantized thermal Hall effect. In Fig.\,\ref{Kappa-1}(d), the crystal with T$_N$=7.6\,K shows a high thermal Hall conductivity $\kappa_{xy}$/T, closer to the half integer quantized value. It is interesting to note that half-integer quantized thermal Hall effect was recently observed for a deformed $\alpha$-RuCl$_3$ crystal with T$_N$=14\,K which exhibits a high thermal Hall resistivity despite a low thermal conductivity\cite{Zhang14K}. These two studies suggest two different ways to observe thermal Hall conductivity at half-integer quantized value: using high quality samples with minimal amount of stacking disorder and high thermal conductivity, or using samples with a high thermal Hall resistivity. Obviously, thermal Hall conductivity can be rather sample dependent and the stacking disorder can have a dramatic effect on it. Despite the sample dependent thermal Hall conductivity, both types of crystals have the similar critical fields for oscillatory features in magnetothermal conductivity and also the same field range in which thermal Hall effect and oscillatory features in thermal conductivity are observed.

The correlation between stacking disorder, magnetic and thermal transport properties developed in this work suggests the following can be \textit{valid criteria when selecting crystals with minimal amount of stacking disorder}: (1) the structure transition temperature measured upon cooling. This can be determined by measuring the temperature dependence of magnetization in the paramagnetic state, or by monitoring the evolution with temperature of an appropriate nuclear reflection by x-ray or neutron diffraction. (2) the magnetic ordering temperature determined by specific heat measurements. In addition to a well defined T$_N$, specific heat measurements can also tell whether extra anomalies exist near T$_N$ (see Fig. S5). We noticed that the anisotropic magnetic susceptibility around T$_N$ is not that sensitive to the small variation of T$_N$ or the presence of multiple magnetic transitions as revealed by specific heat shown in Fig.\,S5. From the characterizations of different crystals in the last decade, crystals with minimal amount of stacking disorder should have a structure transition around 140\,K when being measured upon cooling and one single magnetic anomaly around 7.6\,K in specific heat.

\section{Summary}
In summary, we study the effects of small amount of stacking disorder on the structure, magnetic and thermal transport properties in $\alpha$-RuCl$_3$ single crystals with T$_N$ varying from 6.0\,K to 7.6\,K. The amount of stacking disorder may not be enough to induce magnetic anomalies in the temperature range 10-14\,K, but can still have a dramatic effect on the physical properties. The stacking disorder in as-grown crystals can suppress the structure transition temperature (upon cooling), the Neel temperature, and lattice thermal conductivity. The similar oscillatory field dependent thermal conductivity and plateau like feature in thermal Hall resistivity were observed in all $\alpha$-RuCl$_3$ single crystals studied in this work despite the variation in T$_N$. However, $\alpha$-RuCl$_3$ single crystals with minimal amount of stacking disorder have a higher thermal conductivity, which pushes the thermal Hall conductivity to be closer to the half-integer quantized value. Our results demonstrate a strong correlation between the layer stacking, structure transition, magnetic and thermal transport properties and highlight the importance of interlayer coupling in $\alpha$-RuCl$_3$ despite the weak van der Waals bonding. Our work also suggests well-defined criteria for selecting $\alpha$-RuCl$_3$ crystals with minimal amount of stacking disorder: a structure transition around 140\,K when being measured upon cooling and one single magnetic anomaly around 7.6\,K in specific heat. Moreover, the presence of stacking disorder may be universal in all layered, cleavable magnets. The implications of this work can be extended to other compounds and help resolve and understand their intrinsic properties and the novel quantum phenomena.

\section{Acknowledgment}
JY would thank discussions with Tom Berlijn, Huibo Cao, Hwan Do and Alan Tennant. The authors thank Michael Lance for WDS measurements. HZ, SN, MM, and JY were supported by the U.S. Department of Energy, Office of Science, National Quantum Information Science Research Centers, Quantum Science Center. QZ, DM, AM, HM, and BS were supported by the US Department of Energy,  Office of Science, Basic Energy Sciences, Materials Sciences and Engineering Division. JC and MC were supported by an Early Career project supported by DOE Office of Science FWP ERKCZ55. A portion of this research used resources at the Spallation Neutron Source, a DOE Office of Science User Facility operated by the Oak Ridge National Laboratory.

 This manuscript has been authored by UT-Battelle, LLC, under Contract No.
DE-AC0500OR22725 with the U.S. Department of Energy. The United States
Government retains and the publisher, by accepting the article for publication, acknowledges that the United States Government retains a non-exclusive, paid-up,
irrevocable, world-wide license to publish or reproduce the published form of this
manuscript, or allow others to do so, for the United States Government purposes.
The Department of Energy will provide public access to these results of federally
sponsored research in accordance with the DOE Public Access Plan (http://energy.gov/
downloads/doe-public-access-plan).

\section{references}
%%merlin.mbs apsrev4-1.bst 2010-07-25 4.21a (PWD, AO, DPC) hacked
%Control: key (0)
%Control: author (0) dotless jnrlst
%Control: editor formatted (1) identically to author
%Control: production of article title (0) allowed
%Control: page (1) range
%Control: year (0) verbatim
%Control: production of eprint (0) enabled
%
%\bibliography{RuCl3}

\end{document}